\documentclass[%
 reprint,
superscriptaddress,
,nofootinbib,
 amsmath,amssymb,
 aps,
 pra,
]{revtex4-2}

\usepackage{amsmath}
\usepackage{graphicx}
\usepackage{dcolumn}
\usepackage{bm}
\usepackage{xcolor}
\usepackage[version=4]{mhchem}
\usepackage{xr}
\usepackage{multirow}
\usepackage{array}
\usepackage[title]{appendix}
\usepackage{mathtools}
\usepackage[colorlinks=true,citecolor=blue,linkcolor=red,linktocpage=true]{hyperref}
\usepackage[normalem]{ulem}

\newcommand{\bp}{\mathbf{p}}
\newcommand{\bx}{\mathbf{X}}
\newcommand{\cW}{\mathcal{W}}
\newcommand{\cR}{\mathcal{R}}

\newcommand{\pd}{\text{d}}

\newcolumntype{P}[1]{>{\centering\arraybackslash}p{#1}}

\definecolor{myorange}{RGB}{245,156,74}
\definecolor{mygreen}{RGB}{17,159,87}

\begin{document}

\preprint{APS/123-QED}

\title{Robust topological oscillators govern a tunable phase transition to synchronized circadian rhythms}

\author{Chongbin Zheng}

\affiliation{
 Department of Physics and Astronomy, Rice University, Houston, Texas 77005, USA 
}
\affiliation{
 Center for Theoretical Biological Physics, Rice University, Houston, Texas 77005, USA
}

\author{Peter Thomas}

\affiliation{
Department of Mathematics, Applied Mathematics and Statistics, Case Western Reserve University, Cleveland, Ohio 44106, USA
}

\author{Evelyn Tang}
\thanks{Email: e.tang@rice.edu}

\affiliation{
 Department of Physics and Astronomy, Rice University, Houston, Texas 77005, USA
}
\affiliation{
 Center for Theoretical Biological Physics, Rice University, Houston, Texas 77005, USA
}

\begin{abstract}
While synchronization has been well-studied in deterministic oscillators, most underlying oscillators are stochastic in both natural and man-made systems. Yet, the effects of intrinsic stochasticity remain poorly understood. Here, we develop a new mechanism for synchronizing circadian KaiC molecules that have topologically protected cycles. We find a phase transition to synchronization that depends only on the single-oscillator coherence, across a range of molecular changes that determine this coherence. Examining both mesoscopic and macroscopic numbers relevant for cellular and \textit{in vitro} conditions respectively, we find different scaling properties above and below the phase transition.  Our results shed light on several existing experiments and further predict that external changes can be offset by compensatory changes that improve the single-oscillator coherence -- demonstrating a tunable pathway between stochastic single oscillators and their robust collective rhythms.
\end{abstract}

\maketitle 

\section{Introduction}

Synchronized oscillations are ubiquitous in biological systems \cite{gratz2018synchronization,richard1996sustained,wolf2000effect,richard1993around,singer1999striving,stiefel2016neurons} and have been studied extensively in deterministic models \cite{kuramoto1975,kuramoto1984chemical,pikovsky2001synchronization,acebron2005kuramoto,park2017utility,park2018infinitesimal,kreider2024relation}. However, real biological oscillators are subject to intrinsic stochastic noise, which drives phases apart to prevent synchronization. This motivates the need for models that extend the mechanism of synchronization to stochastic oscillators \cite{neiman1994synchronizationlike,han1999interacting,freund2000analytic,callenbach2002oscillatory,medvedev2010synchronization,zakharova2011analysing,zakharova2013coherence,amro2015phase,deng2016measuring,kreider2025q}. Notably, the onset of synchronization can emerge as a phase transition in the collective dynamics, with the canonical example of the Kuramoto model that couples deterministic oscillators \cite{kuramoto1975,kuramoto1984chemical}. In this framework, synchronization transition arises when the coupling strength exceeds a critical threshold. This framework has been extended to introduce external noise, e.g., in \cite{kreider2025artificial}, but the effects of intrinsic stochastic noise remain largely unexplored. 

Another challenge in the study of synchronization is to move beyond the typical sinusoidal coupling used in the Kuramoto model, which does not directly relate to a physical mechanism. This has motivated the study of more physically grounded couplings by modeling detailed molecular interactions between oscillators \cite{garcia2004modeling,nandi2007effective,to2007molecular,kim2007stochastic}. A particularly interesting system for probing synchronization mechanisms is the KaiABC circadian clock for cyanobacteria. The circadian rhythm is governed by synchronized oscillations in the phosphorylation level of KaiC proteins \cite{cohen2015circadian,chavan2021reconstitution}. This system offers a powerful experimental testbed, since collective oscillations can be reconstituted \textit{in vitro} from just three purified proteins \cite{nakajima2005reconstitution}.

This minimal circadian clock has motivated various deterministic and stochastic models for the emergent oscillations. Although many models have reproduced synchronized KaiC oscillations, the role of the underlying individual molecules remain unclear. Most models describe the population-level dynamics without discussing the role of individual molecules  \cite{rust2007ordered,mori2007elucidating,phong2013robust,lin2014mixtures,mori2018revealing,chew2018high,liu2026underlying}. While some studies do study single KaiC proteins, the resulting behaviors vary from noisy single-KaiC oscillations that dephase quickly \cite{van2007allosteric,paijmans2017thermodynamically} to persistent oscillations powered by intrinsic KaiC ATPase activity \cite{sasai2021mechanism,sasai2022role,sasai2026molecular}. Furthermore, these single-molecule models do not connect to the global phase transition that governs synchronization and often involve complex reaction networks with many parameters. These features can make it difficult to understand and control the resulting collective behavior.

Recently, a minimal model proposed that individual KaiC molecules oscillate in a coherent manner protected by a topological mechanism \cite{zheng2024topological}. The tunability of this minimal model offers an opportunity to investigate the nature of the phase transition to synchronization and shed light on the supporting biophysical mechanisms. Experiments suggest that oscillations arise from negative feedback induced by sequestration of the protein KaiA, which slows down the phosphorylation of leading KaiC molecules \cite{kageyama2006cyanobacterial,brettschneider2010sequestration}. It is hence of interest to add this feedback mechanism to the topological model of KaiC molecules \cite{zheng2024topological}, thereby providing insight into the role that noisy oscillators play in the emergence of synchronization, which is of interest across many natural systems \cite{gonze2002biochemical,gonze2005spontaneous,kim2007stochastic,zhou2008synchronization,pu2020fast}.

\begin{figure*}[htbp!]
    \centering
    \includegraphics[width=16cm, height=14cm]{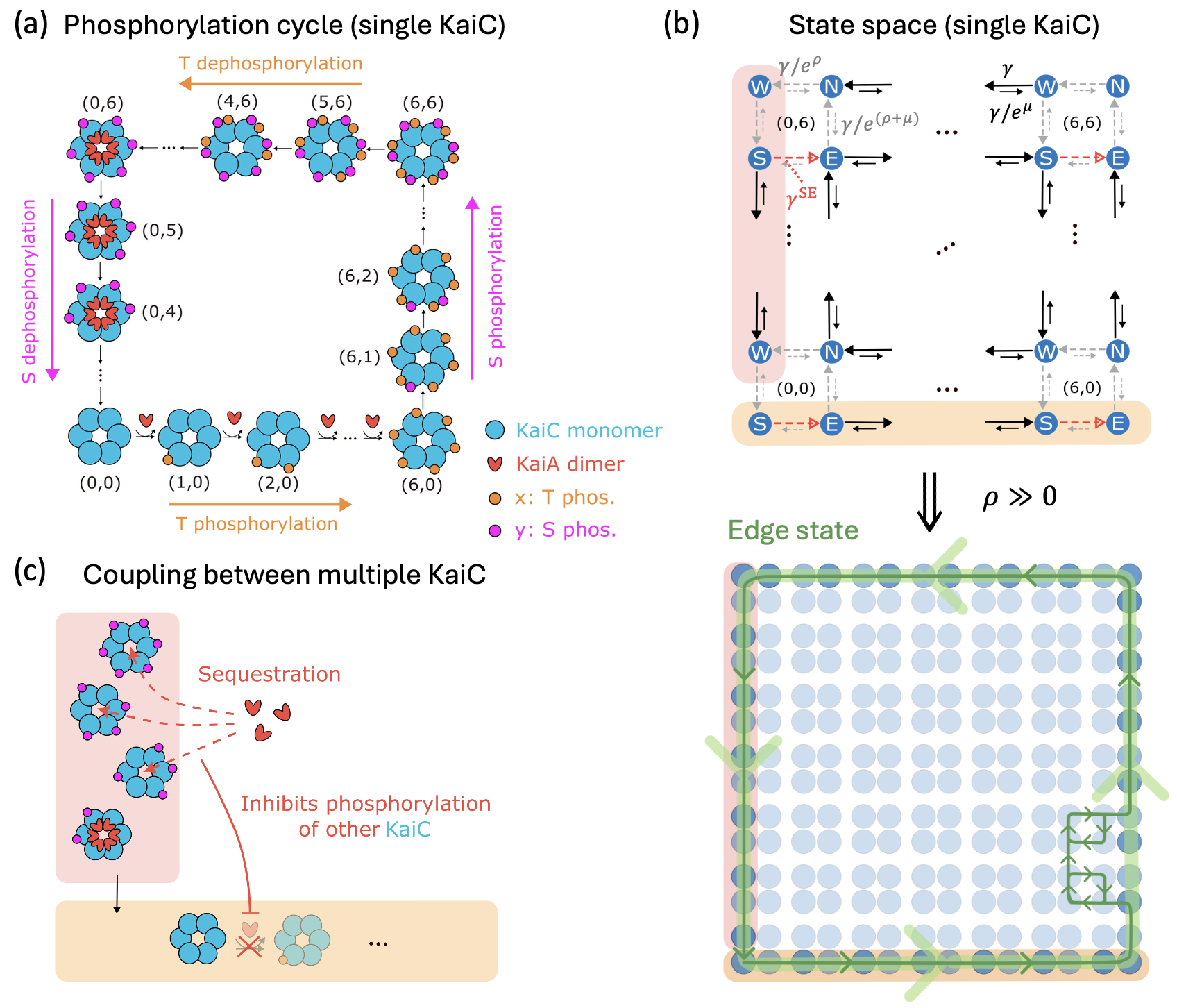}
    \caption{Single-molecule oscillations of KaiC phosphorylation level are synchronized through KaiA sequestration. (a) Individual KaiC hexamers undergo ordered phosphorylation cycles. KaiA dimers (red) promote KaiC phosphorylation and become sequestered during dephosphorylation. $(x,y)$ denotes the number of phosphorylated T and S sites, respectively. (b) Phosphorylation dynamics of a single KaiC molecule can be captured by a topological model with a 2D state space. When $\rho>0$, the system is in a topological phase, where edge currents give rise to autonomous single-molecule oscillations \cite{zheng2024topological}. (c) A population of KaiC oscillators are synchronized through KaiA sequestration. When KaiA is sequestered, phosphorylation slows down for front runners (orange region), allowing the laggards (red region) to catch up.}
    \label{fig_model}
\end{figure*} 

In this paper, we add KaiA sequestration to topologically protected KaiC molecules \cite{zheng2024topological} to find a novel phase transition to synchronization that depends only on the single-oscillator coherence rather than specific biochemical reaction rates. This dependence explains several experimental observations and generates testable predictions, based on how various experimental modifications change single-oscillator coherence. Examining both mesoscopic and macroscopic numbers of molecules, we find different scaling behaviors above and below the phase transition. This model further exhibits robust oscillatory dynamics and oscillation periods across changing KaiA concentrations, which likely arise from the coherent single-molecule dynamics. Our results reveal a new mechanism for synchronization in which collective oscillations depend on the stochastic dynamics of individual oscillators.

\newpage

\begin{figure*}[htbp!]
    \centering
    \includegraphics[width=18cm, height=13.7cm]{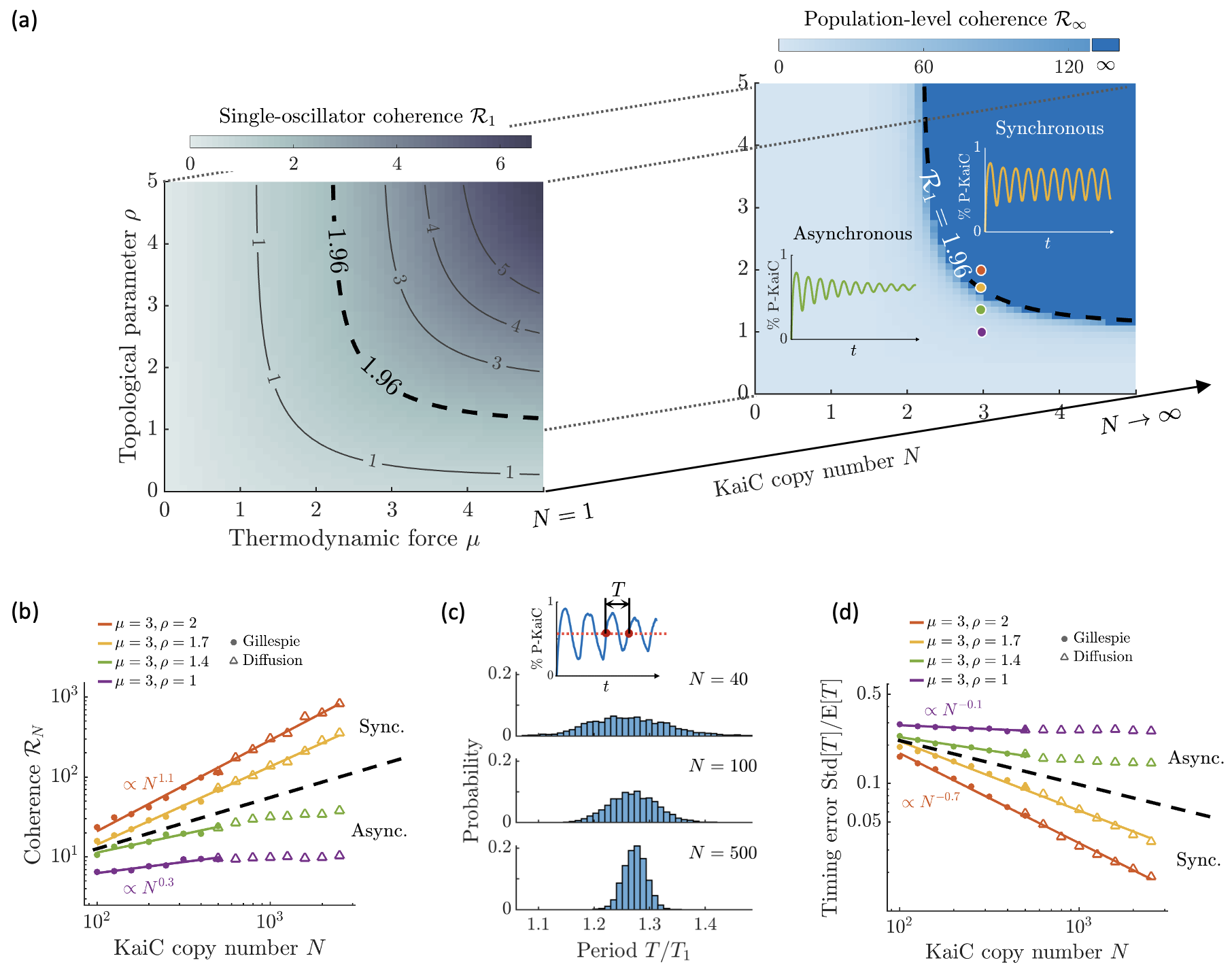}
    \caption{Single-oscillator coherence governs a novel phase transition to synchronization and determines finite-$N$ scaling behavior for collective oscillations. (a) In the deterministic limit $N\rightarrow \infty$, a phase transition to synchronization occurs when the single-oscillator coherence $\cR_1\gtrsim1.96$ (dashed line). The dark blue region corresponds to a synchronous phase with sustained oscillations (e.g., in yellow), while the light blue region corresponds to an asynchronous phase with damped oscillations (in green). (b) Population-level coherence $\cR_N$ grows with $N$ via power laws in the synchronous phase, but stops growing at large $N$ in the asynchronous phase. Circles and triangles represent simulation results from the Gillespie algorithm and the diffusion approximation, respectively. (c) The population-level oscillation period $T$, illustrated on the top, shows smaller timing error at larger $N$. The period is normalized by $T_1$, defined in Sec. \ref{sec_cohr}. (d) Similar to (b), the timing error decreases with $N$ via power laws in the synchronous phase but levels off at large $N$ in the asynchronous phase.}
    \label{fig_phase_transition}
\end{figure*} 

\section{Synchronization of topologically protected KaiC oscillators}
We first characterize the dynamics of a single KaiC hexamer. Experiments show that individual KaiC hexamers undergo ordered phosphorylation cycles at two sites named T and S \cite{rust2007ordered,nishiwaki2007sequential}, as illustrated in Fig.~\ref{fig_model}(a). Following Ref.~\cite{zheng2024topological}, we denote the phosphorylation level of the two sites by $(x,y)$ and model the single-molecule dynamics by discrete stochastic transitions in a two-dimensional state space (Fig.~\ref{fig_model}(b), upper panel). Given phosphorylation level $(x,y)$, letters (N, E, S, W) denote internal states that prime the KaiC hexamer for phosphorylation or dephosphorylation. The system evolves according to the master equation (aka.~Kolmogorov's forward equation \cite{van1983stochastic})
\begin{equation} \label{eqn_master}
    \frac{\pd\bp}{\pd t}=\cW^\text{s}\bp,
\end{equation}
where $\bp\in\mathbb{R}^{M}_+$ encodes the probability distribution over $M=196$ possible states and $\cW^\text{s}$ is the transition-rate matrix for a single KaiC molecule.

The model dynamics is controlled by two parameters. First, the thermodynamic force $\mu$ quantifies the free energy input from ATP hydrolysis (in units of $k_\text{B}T$) \cite{hill1989}, which depends on the ATP/ADP ratio. The system obeys detailed balance at $\mu=0$, while increasing $\mu$ biases each reaction in a preferred direction and drives directed motion in state space. Second, the topological parameter $\rho$ measures the relative rates between phosphorylation/dephosphorylation (solid arrows in Fig.~\ref{fig_model}(b)) and ``internal'' transitions (dashed arrows) that prime these reactions, such as conformational change. Varying $\rho$ tunes the system between a topological phase ($\rho>0$, where phosphorylation is faster) and a trivial phase ($\rho<0$) \cite{zheng2024topological}. In the topological phase, protected edge states give rise to autonomous single-molecule oscillations (Fig.~\ref{fig_model}(b), lower panel). As $\rho$ increases, the edge states become more robust: stochastic trajectories (dark green) are more likely to circulate around the state space boundary rather than wandering into the bulk. The parameters $\mu$ and $\rho$ specify the transition rates for KaiC together with a global scaling factor $\gamma$, as labeled in Fig.~\ref{fig_model}(b).

To synchronize these KaiC oscillators, we couple their dynamics through sequestration of KaiA, a limited resource that promotes KaiC phosphorylation \cite{kageyama2006cyanobacterial,brettschneider2010sequestration}. During dephosphorylation, KaiC hexamers sequester free KaiA dimers and limit their availability to other KaiC molecules \cite{qin2010intermolecular}, as illustrated in Fig.~\ref{fig_model}(c). This negative feedback slows down leading KaiC molecules in the orange region, so that the lagging ones in the red region can catch up. This leads to a narrower distribution of phosphorylation states across the population, which allows the oscillators to synchronize.

In our model, free KaiA dimers promote phosphorylation by speeding up the transitions with rates $\gamma^\text{SE}$ (red dashed arrows in Fig.~\ref{fig_model}(b)). Based on experimental evidence \cite{snijder2017structures,chavan2021reconstitution}, we assume that each KaiC hexamer instantaneously sequesters six KaiA dimers after entering the red region of state space in Fig.~\ref{fig_model}(b), and releases them upon exit. Sequestration of free KaiA slows down the rate $\gamma^\text{SE}$ for every KaiC molecule in the population. Specifically, we let $\gamma^\text{SE}$ decrease linearly with the concentration of sequestered KaiA, denoted by $c_\text{A}^\text{seq}$. This effect modifies the transition-rate matrix $\cW^\text{s}$ by an additional term $c_\text{A}^\text{seq}\cW'$, where $\cW'$ encodes changes to $\gamma^\text{SE}$ (details in Appendix~\ref{appx_rate_eqn}). In the deterministic limit (KaiC copy number $N\rightarrow\infty$), we obtain the rate equation
\begin{equation} \label{eqn_coupled_main}
    \frac{\pd\bx}{\pd t}=\left[\cW^\text{s} + c_\text{A}^\text{seq}(\bx)\cW'\right]\bx,
\end{equation}
where $\bx\in\mathbb{R}^M_+$ denotes the fraction of KaiC molecules in each state. The strength of the coupling depends on the total KaiA concentration $c_\text{A}$. We normalize $c_\text{A}=1$ to the standard experimental condition \textit{in vitro}, where the numbers of KaiA dimers and KaiC hexamers are roughly equal \cite{kageyama2006cyanobacterial,nishiwaki2007sequential}.

\section{Results} \label{sec_results}

\subsection{Single-oscillator coherence determines a phase transition to synchronization} \label{sec_cohr}

To characterize the quality of oscillations in our model, we examine the coherence $\cR_N$ for KaiC copy number $N$ (also called the quality factor) \cite{barato2017coherence,perez2023universal}. 
The quantity measures the number of reliable oscillation cycles before degradation by stochastic noise. For a single oscillator, the oscillation dynamics is governed by the slowest-decaying eigenvalue of the transition-rate matrix $\cW^\text{s}$, denoted by $-\lambda_\text{R}\pm i\lambda_\text{I}$ ($\lambda_\text{R},\lambda_\text{I}>0$). The single-oscillator coherence $\cR_1$ is given by $\cR_1=\lambda_\text{I}/\lambda_\text{R}$ (see details in Appendix~\ref{appx_coherence}). The real part of the eigenvalue specifies the decay time $1/\lambda_\text{R}$ of the oscillations, while the imaginary part provides a natural timescale given by $T_1 = 2\pi/\lambda_\text{I}$, which is close to the period of a single KaiC oscillator \cite{perez2022quantitative}.

The single-oscillator coherence $\cR_1$ increases monotonically with both model parameters $\mu$ and $\rho$ (lower left of Fig.~\ref{fig_phase_transition}(a)). We consider $\mu$ up to 5 $k_\text{B}T$, based on low ATP consumption from experimental measurements of KaiC \cite{terauchi2007atpase}. Meanwhile, we focus on the topological phase $\rho>0$, and take $\rho\in[0,5]$ to explore values both near the phase boundary and deep in the topological phase. We fix the KaiA concentration at the standard condition $c_\text{A}=1$. Notably, the resulting oscillations remain robust, i.e., $\cR_1>1$ \cite{perez2023universal}, across a large range of parameter space. This observation contrasts with earlier models where individual KaiC molecules display low coherence \cite{van2007allosteric,rust2007ordered,mori2007elucidating,phong2013robust,lin2014mixtures,paijmans2017thermodynamically,mori2018revealing,chew2018high}.

Building on these robust oscillators, we examine the population-level coherence $\cR_\infty$ in the deterministic limit ($N\rightarrow\infty$), relevant for \textit{in vitro} experiments with large KaiC copy numbers. $\cR_\infty$ is extracted from oscillations in the percentage of phosphorylated KaiC monomers (\% P-KaiC), described by the rate equation Eq.~(\ref{eqn_coupled_main}) (details in Appendix~\ref{appx_coherence}). Interestingly, we find a phase transition that only depends on the single-oscillator coherence $\cR_1$, not the underlying parameters $\mu$ and $\rho$ that determine $\cR_1$. The transition occurs at $\cR_1\gtrsim1.96$ (dashed lines in Fig.~\ref{fig_phase_transition}(a)). Below the transition (the asynchronous phase), oscillations are damped and $\cR_\infty$ remains finite. Above the transition (the synchronous phase), oscillations become sustained and $\cR_\infty$ diverges. The diverging coherence describes oscillations that no longer decay: the decay time $1/\lambda_\text{R}$ diverges as $N\rightarrow\infty$, while $\lambda_\text{I}$ remains roughly constant, indicating continued oscillations. These results are surprising as the transition depends only on an emergent property of individual oscillators -- their coherence -- rather than specific biochemical reaction rates. The transition is a supercritical Hopf bifurcation, where the oscillation amplitude grows continuously from zero via a square-root scaling law (see analysis in Appendix~\ref{appx_bifurcation}).

\begin{figure*}[htbp!]
	\centering
	\includegraphics[width=17cm, height=6cm]{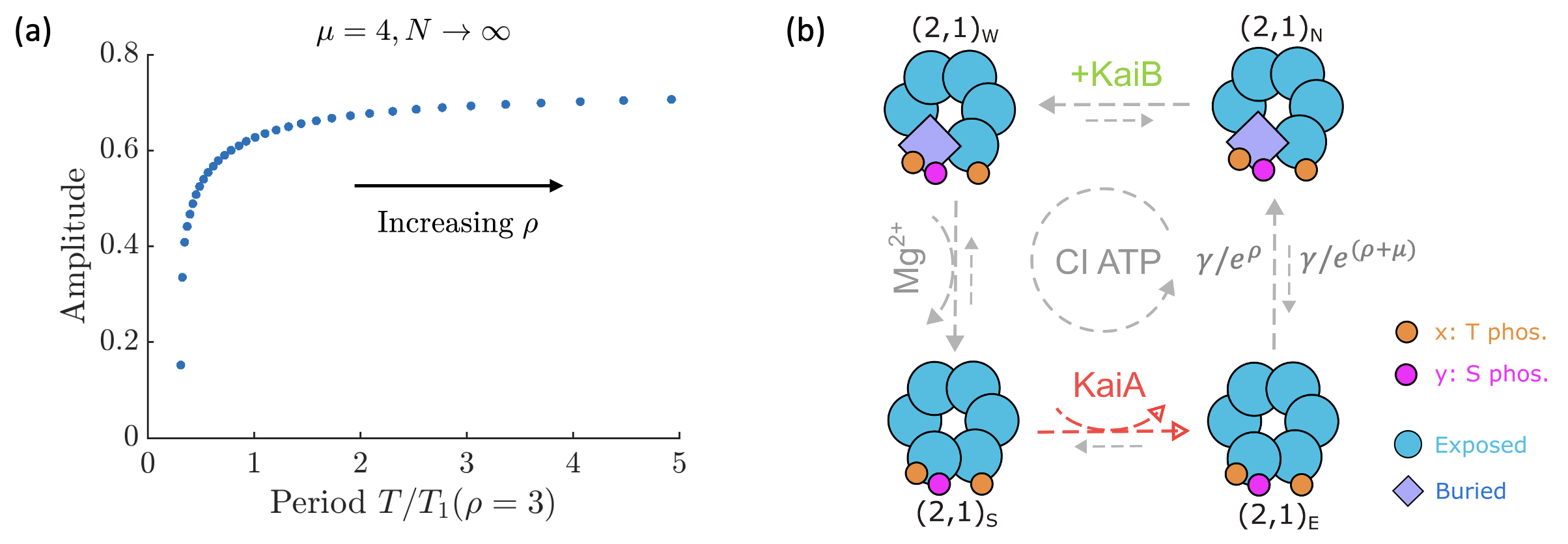}
	\caption{Experimental modifications shape population-level behavior by tuning single-molecule transitions rates. (a) In our model, KaiC mutants with longer population-level period $T$ correspond to a larger topological parameter $\rho$, which leads to larger oscillation amplitudes in \% P-KaiC as experimentally observed \cite{ito2020tuning}. The period is normalized by $T_1$ at $\rho=3$. (b) Internal transitions from Fig.~\ref{fig_model}(b) can be modified by different experimental conditions. KaiB mutants show different binding rates (top), Mg$^{2+}$ concentration regulates A-loop conformational change from buried to exposed (left), KaiA acts catalytically to prime phosphorylation (bottom), while KaiC period mutants modulate CI ATPase activity to set the overall timescale of internal transitions (center).}
    \label{fig_exp}
\end{figure*} 

\subsection{Synchronous and asynchronous phases exhibit distinct scaling behavior with $N$}

The above results connect single-molecule dynamics to collective oscillations in the limit $N\rightarrow\infty$. For oscillations \textit{in vivo}, however, only thousands of KaiC molecules are present \cite{kitayama2003kaib,chew2018high} and stochastic fluctuations remain significant. To characterize oscillations at finite $N$, we perform stochastic simulations and extract the coherence $\cR_N$ numerically (details in Appendix~\ref{appx_coherence}). For $N\leq500$, we obtain $\cR_N$ using the Gillespie algorithm \cite{gillespie1977exact} (circles in Fig.~\ref{fig_phase_transition}(b)). At larger $N$, we instead use a diffusion approximation \cite{goldwyn2011and,orio2012simple,pu2020fast} for computational efficiency, shown by triangles in Fig.~\ref{fig_phase_transition}(b) (detailed derivations in Appendix~\ref{appx_diffusion}). $\cR_N$ displays different scaling behavior with $N$ depending on the phase of the system. In the synchronous phase, $\cR_N$ grows with $N$ via power laws and is expected to diverge at $N\rightarrow\infty$. In the asynchronous phase, $\cR_N$ initially grows via power laws but eventually plateaus as $N$ increases.

While the two phases can be distinguished by different scaling behavior in $\cR_N$, a quantity that is easier to measure experimentally is the timing error. It quantifies the variability of the population-level period $T$ and is defined as $\text{Std}[T]/\text{E}[T]$ \cite{chew2018high}. In our simulations, we extract $T$ by recording when \% P-KaiC increases past a threshold, shown on the top of Fig.~\ref{fig_phase_transition}(c) (details in Appendix~\ref{appx_timing_error}). The resulting distributions at different $N$ are shown in Fig.~\ref{fig_phase_transition}(c), where we take $\mu=\rho=5$. The timing error displays similar scaling behavior as $\cR_N$: it decreases with $N$ via power laws in the synchronous phase but eventually plateaus in the asynchronous phase (Fig.~\ref{fig_phase_transition}(d)). These different scaling behavior provide an experimentally accessible way to detect the different phases at finite $N$.

\subsection{The phase transition accounts for several existing experimental observations} \label{sec_exp}

This phase transition framework sheds light on several experimental observations. First, longer periods in our model are associated with larger oscillation amplitudes, as shown in Fig.~\ref{fig_exp}(a). This is because longer periods correspond to a larger topological parameter $\rho$, which enhances the single-oscillator coherence $\cR_1$. A higher $\cR_1$ allows the population to oscillate more in phase with each other, yielding a larger oscillation amplitude (see details in Appendix \ref{appx_R1_amp}). Second, stronger non-equilibrium driving also produces larger oscillation amplitudes in our model, which can be similarly explained by a larger $\cR_1$. Third, large changes to individual biochemical reaction rates (e.g., those in Fig.~\ref{fig_exp}(b)) can disrupt sustained population-level oscillations. The oscillations are lost because these changes degrade $\cR_1$, which drive the system to the asynchronous phase. In each case above, changes in collective behavior directly follow from changes in $\cR_1$, without requiring detailed simulations of the experimental modifications as in other models. 

The first result, where a longer period leads to a larger amplitude, has been observed for several KaiC mutants with different periods \cite{ito2020tuning}. Our model predicts this because the different mutants correspond to different topological parameter $\rho$ and thus different $\cR_1$. The longer-period mutants have lower CI ATPase activity, which slows down the internal transitions shown in Fig.~\ref{fig_exp}(b) \cite{terauchi2007atpase,tseng2017structural,mukaiyama2018conformational}. Since the internal transition rates scale as $1/e^\rho$, longer periods lead to larger $\rho$ and larger $\cR_1$, which produces a larger amplitude. The amplitude grows more rapidly at shorter periods (Fig.~\ref{fig_exp}(a)), which also agrees with experimental observations \cite{ito2020tuning}. Similar analysis also explains the second result of larger amplitudes at stronger non-equilibrium driving, observed in experiments with different ATP/ADP ratios \cite{phong2013robust}. In our model, a higher ATP/ADP ratio sets a stronger thermodynamic force $\mu$, which also increases $\cR_1$ and the amplitude.

Experiments have further demonstrated the third result, where sustained oscillations are lost under disruptive changes such as extreme Mg$^{2+}$ levels \cite{jeong2019magnesium} or KaiB mutants \cite{hitomi2005tetrameric,chang2015protein}. These modifications act on individual reaction rates that have been mapped to specific transitions in our model \cite{zheng2024topological}, as illustrated in Fig.~\ref{fig_exp}(b). KaiB binding to KaiC corresponds to the $(x,y)_\text{N}\rightarrow (x,y)_\text{W}$ transition, which becomes faster for the KaiB mutants in \cite{hitomi2005tetrameric,chang2015protein}. Mg$^{2+}$ ions enable conformational changes of KaiC A-loops from exposed (blue circles) to buried (purple squares) \cite{jeong2019magnesium}, which promotes the $(x,y)_\text{W}\rightarrow (x,y)_\text{S}$ transition. Large changes to these rates degrade $\cR_1$ below the critical threshold, pushing the system into the asynchronous phase. Together, these three results illustrate the strength of our framework: the effects of disparate experimental conditions can be understood in a unified way by analyzing changes to $\cR_1$, which also provides a direct connection between single-molecule dynamics and collective behavior.

\begin{figure*}[htbp!]
	\centering
	\includegraphics[width=17cm, height=12.3cm]{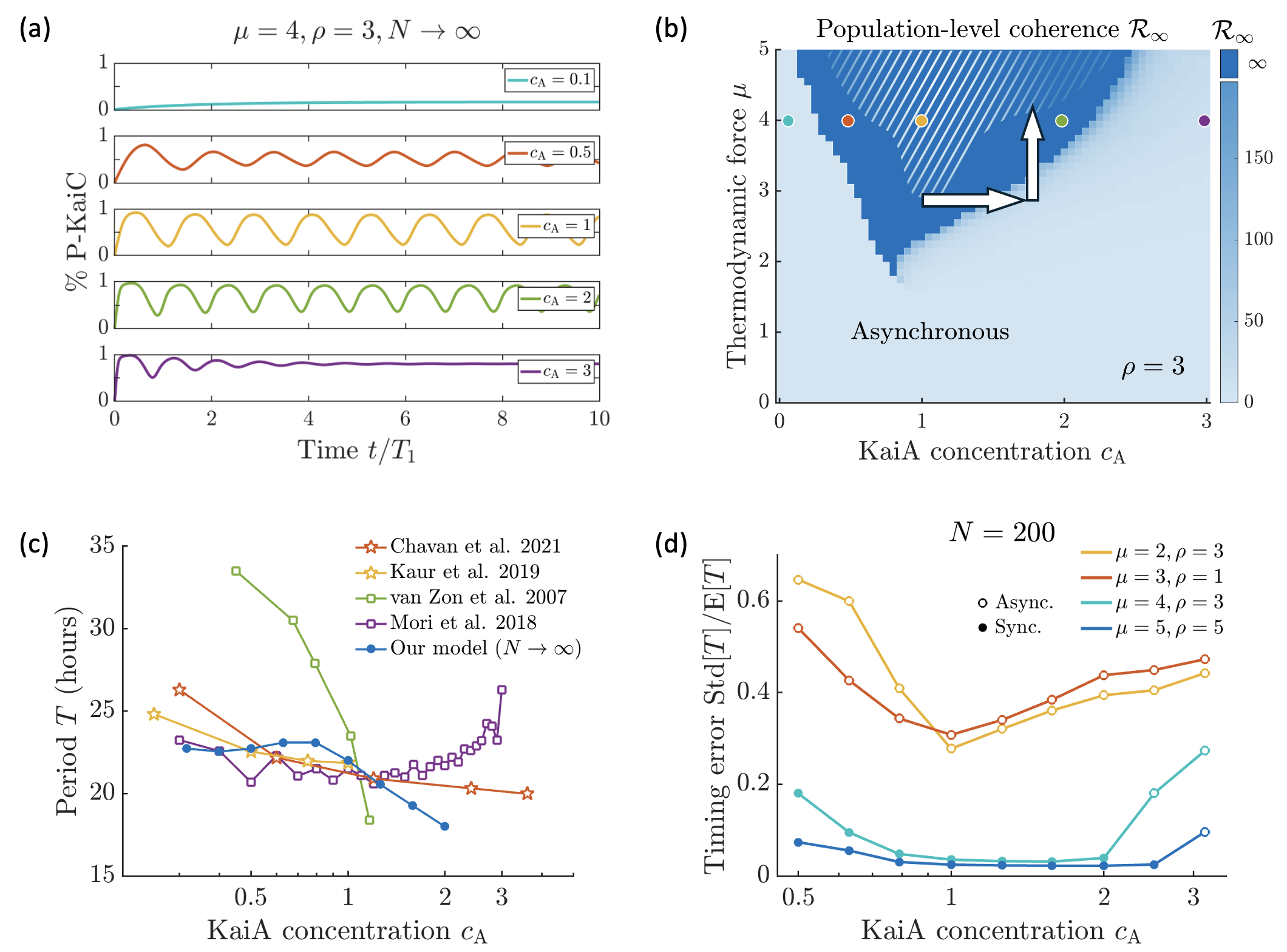}
	\caption{KaiC oscillations and collective period are robust under changing KaiA concentration $c_\text{A}$. (a) Sustained oscillations in \% P-KaiC persist over a wide range of $c_\text{A}$. Different colors correspond to different $c_\text{A}$ shown in (b). (b) The range of $c_A$ supporting sustained oscillations expands with the thermodynamic force $\mu$. Oscillations are lost when $c_\text{A}$ is too high or too low, but can be restored by increasing $\mu$ (white arrows). The hatched area indicate the parameter range for which sustained KaiC oscillations are observed experimentally \cite{kageyama2006cyanobacterial,terauchi2007atpase,phong2013robust,chavan2021reconstitution}. (c) The oscillation period in our model (blue circles) is robust under changing $c_\text{A}$, consistent with experimental data (stars) \cite{chavan2021reconstitution,kaur2019cika}. In comparison, the period from other models (squares) either shows a stronger dependence on $c_\text{A}$ \cite{van2007allosteric} (green) or increases at large $c_\text{A}$ instead \cite{mori2018revealing} (purple). (d) The timing error remains low over a broad range of $c_\text{A}$ in the synchronous phase (filled circles), and increases rapidly when $c_\text{A}$ deviates from one in the asynchronous phase (open circles).}
    \label{fig_KaiA}
\end{figure*} 

Because the phase transition only depends on $\cR_1$, we further predict that disruptions to synchronized oscillations can be offset by other changes that restore $\cR_1$ above the critical threshold. For example, sustained oscillations disrupted by KaiB mutants or low Mg$^{2+}$ concentration may be recovered by increasing the ATP/ADP ratio or using longer-period KaiC mutants. Notably, these restoring changes do not directly reverse the disrupted processes -- adding ATP does not slow down KaiB binding, and KaiC mutants do not restore Mg$^{2+}$-dependent conformational changes. Instead, they raise $\cR_1$ back above the critical value by acting on unrelated biochemical processes. Such compensatory effects are a distinctive prediction of our model. In previous models \cite{van2007allosteric,rust2007ordered,mori2007elucidating,phong2013robust,lin2014mixtures,paijmans2017thermodynamically,liu2026underlying,mori2018revealing,chew2018high,sasai2021mechanism,sasai2022role,sasai2026molecular}, the disruptions above would affect key mechanistic steps that enable population-level oscillations, which are unlikely to be rescued by changes to unrelated rates.

\subsection{Synchronized KaiC oscillations and their period remain robust under changing KaiA concentration} \label{sec_KaiA}

The coherent single-molecule dynamics further lead to robust population-level oscillations across different KaiA concentration $c_\text{A}$, even though $c_\text{A}$ changes the coupling between KaiC oscillators. In our model, sustained oscillations persist over a wide range of $c_\text{A}$ (see Fig.~\ref{fig_KaiA}(a)), consistent with experimental observations \cite{kageyama2006cyanobacterial,chavan2021reconstitution}. When $c_\text{A}$ is too high, oscillations are lost via a supercritical Hopf bifurcation. When $c_\text{A}$ is too low, the system instead undergoes a subcritical Hopf bifurcation followed by a saddle-node bifurcation of limit cycles, where the oscillation amplitude drops to zero abruptly (details in Appendix~\ref{appx_bifurcation}). 

The range of $c_\text{A}$ supporting sustained oscillations expands with higher single-oscillator coherence, as illustrated in Fig.~\ref{fig_KaiA}(b) for increasing thermodynamic force $\mu$. The hatched region indicates the estimated parameter range for past experiments where sustained KaiC oscillations have been observed \cite{kageyama2006cyanobacterial,terauchi2007atpase,phong2013robust,chavan2021reconstitution} (see Appendix~\ref{appx_estimate}). As a result, KaiC synchronization becomes more robust to changing coupling strength as single-molecule oscillations improve. We thus predict that disruptions in population-level oscillations from changing $c_\text{A}$ can be offset by increasing the ATP/ADP ratio that raises $\mu$ (white arrows in Fig.~\ref{fig_KaiA}(b)). This extends the compensatory mechanism in Sec. \ref{sec_exp}: increased single-molecule coherence can compensate for disruptive changes in coupling strength as well as in single-molecule transition rates.

Within the synchronous phase, our model yields a robust oscillation period $T$ under changing $c_\text{A}$, as observed by \textit{in vitro} experiments \cite{chavan2021reconstitution,kaur2019cika} (red and yellow stars in Fig.~\ref{fig_KaiA}(c)). Blue circles show our model results in the deterministic limit, with $\mu=4,\rho=3$ and $T$ normalized to 22 hours at $c_\text{A}=1$. The stability of the period arises from coherent single-molecule oscillations, whose intrinsic period $T_1$ anchors the timescale for population-level oscillations. Even though varying $c_\text{A}$ can modulate the phosphorylation and sequestration dynamics, it only perturbs $T$ away from this baseline. 

In contrast, no such timescale can be defined in previous models with incoherent single-KaiC oscillations \cite{van2007allosteric,paijmans2017thermodynamically}. The period instead emerges from KaiA-mediated coupling, and depends more strongly on $c_\text{A}$ as seen in van Zon et al. \cite{van2007allosteric} (green squares in Fig.~\ref{fig_KaiA}(c)). To offset the rapid decrease in $T$, Mori et al. \cite{mori2018revealing} incorporate a longer sequestration time for larger $c_\text{A}$. This yields a more robust period but an increasing trend at large $c_\text{A}$ (purple squares). These difficulties highlight the importance of coherent single-molecule oscillations, which provide a natural mechanism for stabilizing the period.

Coherent single-molecule oscillations further contribute to a robust timing error under changing $c_\text{A}$. In the synchronous phase (filled circles in Fig.~\ref{fig_KaiA}(d)), the timing error remains low over a wide range of $c_\text{A}$. It only starts to increase at lower $c_\text{A}$, where the oscillation amplitude shrinks and becomes comparable to stochastic fluctuations. This robustness likely arises from coherent single-molecule dynamics, as parameters with higher $\cR_1$ (dark blue) maintain low timing error over a wider range of $c_\text{A}$. In the asynchronous phase (open circles), however, the timing error increases rapidly as $c_\text{A}$ deviates from the standard condition $c_\text{A}=1$. 

\section{Discussion}
In this study, we have developed a population-level model for synchronized oscillations of KaiC phosphorylation. Our model identifies a phase transition to synchronization among coupled stochastic KaiC oscillators, which has not been analyzed in previous models of KaiC \cite{van2007allosteric,rust2007ordered,mori2007elucidating,phong2013robust,lin2014mixtures,paijmans2017thermodynamically,liu2026underlying,mori2018revealing,chew2018high,sasai2021mechanism,sasai2022role,sasai2026molecular}. Interestingly, the transition only depends on the single-oscillator coherence $\cR_1$, providing a direct connection between single-molecule stochastic dynamics and population-level oscillations. It further sheds light on the effects of diverse experimental modifications, which can be understood through how they modify a single emergent property, $\cR_1$, without detailed population-level simulations that account for each modification. We further characterize the nature of the phase transition: it is generally a supercritical Hopf bifurcation, but shows up as a subcritical Hopf bifurcation along with a saddle-node bifurcation of limit cycles at low KaiA concentration.

Our model further yields several distinctive predictions that can be tested experimentally. Because the phase transition only depends on single-oscillator coherence rather than specific reaction rates, we predict that experimental modifications that disrupt population-level oscillations may be compensated by other modifications that increase single-oscillator coherence. For example, damped oscillations resulting from KaiB mutants or low Mg$^{2+}$ levels may be restored by increasing the ATP/ADP ratio or using KaiC mutants with longer periods. Similar compensatory changes may also restore sustained oscillations for extreme levels of KaiA concentration, which disrupt the coupling between KaiC molecules. Our model also predicts that the timing error for synchronized oscillations remains robust under changing KaiA concentration, provided that the oscillation amplitude is large compared to stochastic fluctuations. Finally, when KaiA concentration is reduced, we predict that the oscillation amplitude drops discontinuously to zero due to a subcritical Hopf bifurcation (see Sec.~\ref{sec_KaiA}). This contrasts with a previous model from van Zon et al. \cite{van2007allosteric}, where amplitudes are expected to change continuously through a supercritical Hopf bifurcation.

Overall, our results demonstrate how robust collective oscillations can arise from coherent molecular oscillators, even under limited coupling or changing external conditions. By identifying single-oscillator coherence as a key parameter governing synchronization, we show that mechanisms that maintain molecular-scale coherence (e.g., topological protection \cite{zheng2024topological}) can play an important role in sustaining population-level oscillations. Beyond the KaiABC system, modifying single-molecule properties may provide a general strategy for designing and controlling reliable biochemical oscillators. More broadly, the methods developed here provide a framework for connecting single-molecule stochastic dynamics to emergent collective behavior, which may provide insight on robust dynamics in other stochastic biological systems, such as neuronal \cite{buzsaki2006rhythms} or metabolic \cite{dano1999sustained,wolf2000effect} networks.

\section*{Acknowledgments}
CZ and ET acknowledge support from the NSF Center for Theoretical Biological Physics (PHY2019745), the NSF CAREER Award (DMR-2238667),
and the CZI Theory Institute Without Walls. PT acknowledges support by NSF grant DMS-2052109 and the Oberlin College Department of Mathematics. The authors declare that they have no competing interests. All data needed to evaluate the conclusions in the paper are present in the paper and the Supplementary Materials.

\begin{appendices} 
\section{The rate equation} \label{appx_rate_eqn}
In this section, we provide a more detailed account of the deterministic rate equation Eq.~(\ref{eqn_coupled_main}), which governs the dynamics of a coupled population of KaiC oscillators.Let the population vector $\bx(t)=(X_1(t),X_2(t),...,X_M(t))^\intercal$ denote the fraction of molecules 
in each of the $M$ states. For an infinite population of independent KaiC oscillators, $\bx(t)$ evolves according to
\begin{equation} \label{eqn_rate_single}
    \frac{\pd\bx}{\pd t}=\cW^\text{s}\bx,
\end{equation}
where $\cW^\text{s}_{ij}$ encodes the transition rate from state $j$ to $i$. 
This equation is analogous to the master equation for a single KaiC molecule (Eq.~(\ref{eqn_master}) in the main text):
\begin{equation} \label{eqn_master_single}
    \frac{\pd\bp}{\pd t}=\cW^\text{s}\bp,
\end{equation}
where $\bp(t)$ is the probability distribution over different states. 
In the deterministic limit $N\rightarrow\infty$, fluctuations are averaged out and the population vector $\bx(t)$ evolves in the same way as the probability distribution $\bp(t)$.

We next incorporate the effects of KaiA sequestration, which modifies the KaiA-promoted transition rate $\gamma^\text{SE}$ in $\cW^\text{s}$. 
Let $c_\text{A}^\text{free}$, $c_\text{A}^\text{seq}$, and $c_\text{A}$ denote the free, sequestered, and total KaiA concentration, respectively. 
The total KaiA concentration $c_\text{A}$ is kept fixed, while sequestration reduces the amount of free KaiA and thereby lowers $\gamma^\text{SE}$. 
We model this effect by letting $\gamma^\text{SE}$ vary linearly with $c_\text{A}^\text{free}$, which in turn depends on the population vector $\bx(t)$:
\begin{align} \label{eqn_gammain_SE}
\begin{split}
    \gamma^\text{SE}(\bx) &= c_\text{A}^\text{free}(\bx)(\gamma/e^{\rho}-\gamma/e^{\mu+\rho}) + \gamma/e^{\mu+\rho} \\
    &= \left[c_\text{A}-c_\text{A}^\text{seq}(\bx)\right](\gamma/e^{\rho}-\gamma/e^{\mu+\rho}) + \gamma/e^{\mu+\rho}.
\end{split}
\end{align}
The rate is defined such that $\gamma^\text{SE}=\gamma/e^{\rho}$ at $c_\text{A}^\text{free}=1$ and $\gamma^\text{SE}=\gamma/e^{\mu+\rho}$ (same as its reverse rate) when all KaiA is sequestered at $c_\text{A}^\text{free}=0$. 
We normalize the total KaiA concentration so that $c_\text{A}=1$ corresponds to the standard condition for \textit{in vitro} experiments \cite{kageyama2006cyanobacterial,nishiwaki2007sequential}. 

At the standard condition, the KaiA dimer concentration approximately equals the KaiC hexamer concentration \cite{kageyama2006cyanobacterial,nishiwaki2007sequential}. 
Here we set them to be exactly equal, so that $c_\text{A}$ can be interpreted as the total KaiA concentration relative to that of KaiC. 
Thus $c_\text{A}^\text{seq}$ can  be obtained from the distribution of KaiC via
\begin{equation}
    c_\text{A}^\text{seq}(\bx) = \text{min}(\mathbf{v}_\text{A}^\intercal\bx,c_\text{A}),
\end{equation}
where the vector $\mathbf{v}_\text{A}$ encodes the number of sequestered KaiA dimers in each KaiC state. 
We take $\mathbf{v}_{\text{A},i}=6$ for states in the red region of Fig.~\ref{fig_model}(b) and $\mathbf{v}_{\text{A},i}=0$ otherwise. At finite KaiC copy number $N$, the concentration $c_\text{A}$ translates to a discrete count of KaiA dimers: a population of $N$ 
KaiC hexamers corresponds to $c_\text{A}N$ KaiA dimers.

Based on Eq.~(\ref{eqn_gammain_SE}), the effects of sequestration can be expressed compactly as an additional coupling term $c_\text{A}^\text{seq}(\bx)\cW'$ that modifies the single-molecule transition matrix $\cW^\text{s}$. We thus obtain the deterministic rate equation for the coupled population:
\begin{equation} \label{eqn_coupled}
    \frac{\pd\bx}{\pd t}=\left[\cW^\text{s} + c_\text{A}^\text{seq}(\bx)\cW'\right]\bx.
\end{equation}
The matrix $\cW'$ encodes changes to the transition rate $\gamma^\text{SE}$. Specifically, $\cW'_{ij}=-(\gamma/e^\rho-\gamma/e^{\mu+\rho})$ for the transitions $(x,y)_\text{S}\rightarrow(x,y)_\text{E}$ and $\cW'_{ij}=0$ otherwise. 

We numerically solve Eq.~(\ref{eqn_coupled}) to obtain the population dynamics and the coherence in Fig.~\ref{fig_phase_transition}(a) of the main text. The solution $\bx(t)$ gives the distribution of KaiC molecules across different possible states. To convert $\bx(t)$ to the fraction of phosphorylated monomers, or \% P-KaiC for short (denoted by $f(t)$), we compute the inner product $f(t)=\mathbf{f}^\intercal\bx(t)$ at each time step, where the vector $\mathbf{f}=(f_1,f_2,...,f_n)^\intercal$ denotes the \% P-KaiC corresponding to each state. For a state $i$ with phosphorylation level $(x,y)$, we take $f_i=\frac{1}{6}(x+y-\frac{xy}{6})$. This expression corresponds to the expected fraction of monomers with at least one phosphorylated site, assuming that the $x$ phosphorylated T sites and $y$ phosphorylated S sites are randomly distributed among the six monomers.

\section{Coherence} \label{appx_coherence}
In this section, we define coherence and describe how we extract it from \% P-KaiC oscillations. 
We begin with a single KaiC molecule and then extend the definition to a population. The single-molecule dynamics is described by the master equation Eq.~(\ref{eqn_master_single}). 
It has the general solution $\bp(t)=\text{exp}\left(\cW^\text{s}t\right)\bp(0)$, which can be decomposed into eigenmodes of $\cW^\text{s}$. At long times, the dynamics is dominated by the slowest-decaying eigenvalue, i.e., the eigenvalue with the least negative real part \cite{thomas2014asymptotic}. 
For KaiC oscillators, the eigenvalue 
appears as a complex conjugate pair, which we denote by $\lambda_1=-\lambda_\text{R}\pm i\lambda_\text{I}$ ($\lambda_\text{R}, \lambda_\text{I} \geq0$). The imaginary part $\lambda_\text{I}$ sets the oscillation period for the probability distribution $\bp(t)$, which provides a natural timescale $T_1=2\pi/\lambda_\text{I}$ for the system dynamics. We use the period $T_1$ to normalize the timescale of population-level oscillations in Fig.~\ref{fig_phase_transition}(c) and Fig.~\ref{fig_KaiA}(a). Using other definitions of the period (e.g. the mean--return--time period \cite{schwabedal2013phase}) do not qualitatively affect our conclusions.

The slowest-decaying eigenvalue $\lambda_1$ also governs oscillations of \% P-KaiC, averaged over different stochastic trajectories. 
Using the vector $\mathbf{f}$ defined in Appendix~\ref{appx_rate_eqn}, the expectation value of \% P-KaiC follows
\begin{equation}
    \langle f(t)\rangle=\mathbf{f}^\intercal \bp(t) = \mathbf{f}^\intercal\text{exp}\left(\cW^\text{s}t\right)\bp(0),
\end{equation}
which exhibits sinusoidal oscillations with an exponentially decaying envelope. 
The oscillation period is again given by $2\pi/\lambda_\text{I}$, while the exponential envelope has a characteristic decay time $1/\lambda_\text{R}$ \cite{thomas2014asymptotic,barato2017coherence}. 
Following Ref.~\cite{barato2017coherence}, we thus define the single-oscillator coherence (also called the oscillation's quality factor \cite{perez2023universal}) by 
\begin{equation}
    \cR_1 = \frac{\lambda_\text{I}}{\lambda_\text{R}},
\end{equation}
which corresponds to the number of coherent oscillations in \% P-KaiC before relexation to the steady state. From this definition, we obtain $\cR_1$ in the lower left of Fig.~\ref{fig_phase_transition}(a) by computing the eigenvalues of $\cW^\text{s}$ across different $\mu$ and $\rho$.

For a coupled population of KaiC molecules, \% P-KaiC may also go through damped oscillations similar to the single-molecule case. However, the eigenvalue $\lambda_1$ cannot be easily obtained, as the transitions matrix changes over time due to KaiA sequestration. We thus extract the population-level coherence from the time trace of \% P-KaiC oscillations. In the deterministic limit $N\rightarrow\infty$, we numerically solve the rate equation for the population (Eq.~(\ref{eqn_coupled})), where all molecules are initialized in the state $(0,0)_\text{S}$. We then identify the peaks of the oscillations and fit them to a decaying exponential of the form $\beta_1+\beta_2e^{-\beta_3 t}$ ($\beta_1,\beta_2,\beta_3\in\mathbb{R}$), using the nlinfit function in MATLAB. The first two peaks are discarded to eliminate the effects of transient behavior from the initial condition. We also compute the average time interval between successive peaks, denoted by $\bar{T}$. The coherence is then given by 
\begin{equation} \label{eqn_coherence_num}
    \cR_\infty=\frac{2\pi}{\beta_3 \bar{T}},
\end{equation}
which is shown on the upper right of Fig.~\ref{fig_phase_transition}(a). When the oscillations do not decay, the coherence is considered to diverge, i.e., $\cR_\infty\rightarrow\infty$. 

For finite $N$, because \% P-KaiC oscillations become noisy, we generate at least 1000 stochastic trajectories and study their average at each $N$. Again, all oscillators are initialized at state $(0,0)_\text{S}$ for each simulation. The averaged trajectories are  smoothed using the Savitsky-Golay filter with third-order polynomial fitting and a window size of 41 time points. We then extract the peaks from the smoothed trajectories, obtain $\beta_3$ and $\bar{T}$ as above, and calculate $\cR_N$ using Eq.~(\ref{eqn_coherence_num}). The same procedure is performed for both Gillespie simulations and the diffusion approximation.

\begin{figure*}[ht!]
	\centering
	\includegraphics[width=16.5cm, height=12cm]{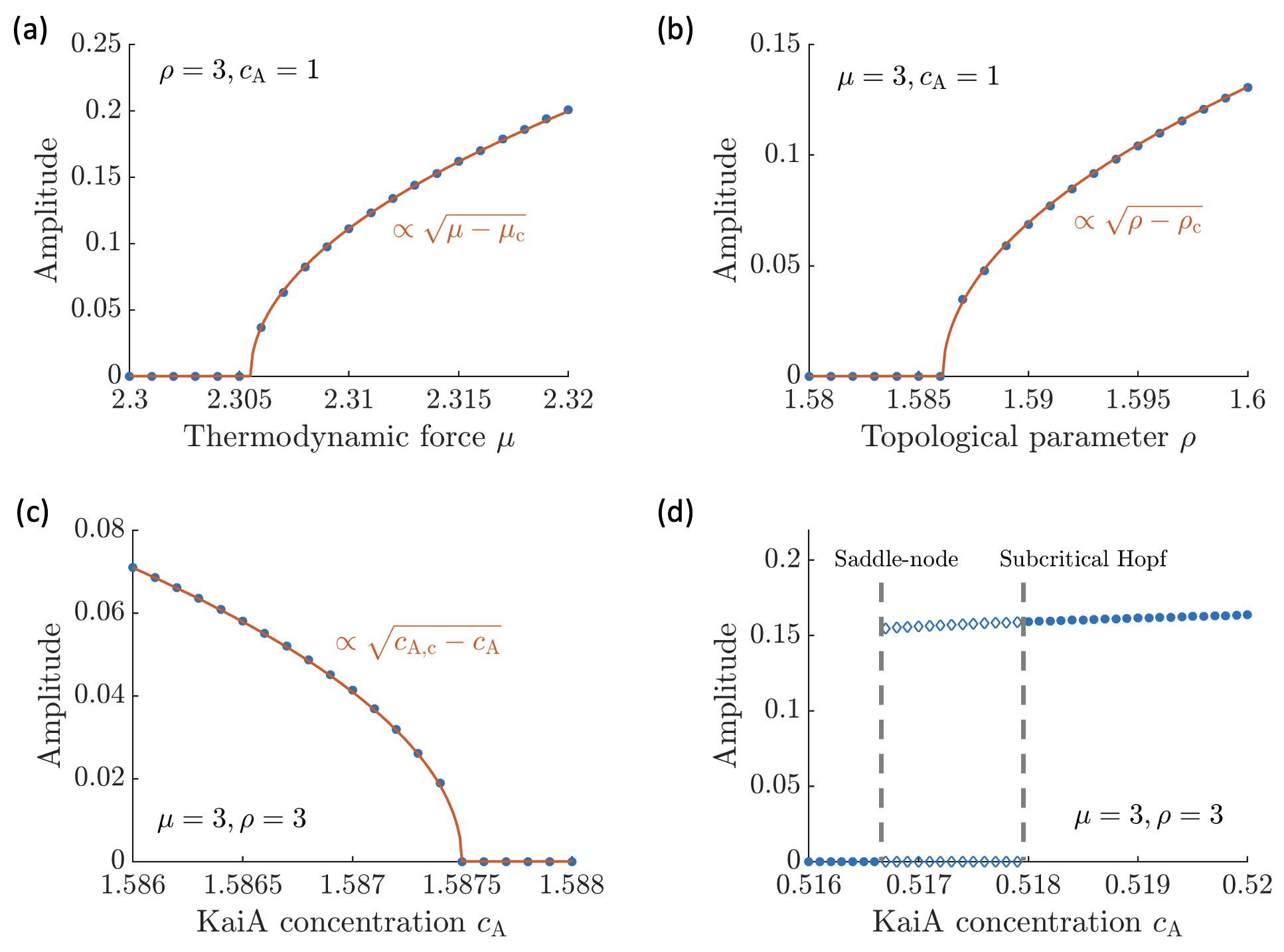}
	\caption{Oscillation amplitude near phase boundaries reveal qualitatively different bifurcations. (a) At the standard KaiA concentration $c_\text{A}=1$ and fixed $\rho$, the oscillation amplitude of \% P-KaiC grows as $\sqrt{\mu-\mu_\text{c}}$ when the thermodynamic force $\mu$ exceeds some critical value $\mu_\text{c}$, a hallmark of a supercritical Hopf bifurcation. (b) The same square root scaling $\sqrt{\rho-\rho_\text{c}}$ is observed when tuning the topological parameter $\rho$ at $c_\text{A}=1$ and fixed $\mu$. (c) At high $c_\text{A}$ and fixed $(\mu,\rho)$, the amplitude decreases as $\sqrt{c_{\text{A},\text{c}}-c_\text{A}}$, indicating another supercritical Hopf bifurcation. (d) At low $c_\text{A}$, the transition is qualitatively different. The system first enters a bistable regime with a stable limit cycle (nonzero amplitude) and a coexisting stable fixed point (zero amplitude), shown by the two branches of diamonds. As $c_\text{A}$ decreases further, the stable limit cycle annihilates with an unstable limit cycle in a saddle–node bifurcation of limit cycles, and only the stable fixed point remains. In all panels, blue markers denote numerical results from our model while red curves in panels (a)-(c) show square-root fits.}
    \label{fig_amplitude}
\end{figure*} 

\section{Phase transition to synchronization} \label{appx_bifurcation}
In this section, we analyze the bifurcations in Fig.~\ref{fig_phase_transition}(a) and Fig.~\ref{fig_KaiA}(b) that result in the transition to synchronization. To identify the types of bifurcations involved, we examine the eigenvalues of the Jacobian at the steady state $\bx^\text{ss}$ of Eq.~(\ref{eqn_coupled}). First, we solve for $\bx^\text{ss}$ numerically by setting $\frac{\pd\bx}{\pd t}=0$. Linearizing the dynamics around the steady state gives the Jacobian:
\begin{equation}
    J = \cW^\text{s} + (\mathbf{v}_\text{A}^\intercal\, \bx^\text{ss})\cW' + \cW' (\bx^\text{ss}\, \mathbf{v}_\text{A}^\intercal).
\end{equation}
This Jacobian always has a zero eigenvalue, because each of its columns sums to 0. This mode reflects conservation of probability and does not affect the oscillatory dynamics, so we exclude it from further analysis. Among the remaining modes, the slowest decaying ones appear as a complex conjugate pair. 
As the system parameters ($\mu,\rho$, or $c_\text{A}$) are tuned to the synchronous phase, the pair of eigenvalues moves across the imaginary axis, where their real parts become positive. 
As a result, the fixed point $\bx^\text{ss}$ becomes unstable and a stable limit cycle emerges, indicating a  Hopf bifurcation that marks the onset of synchronous oscillations.

To further determine the nature of the Hopf bifurcation, we examine how the oscillation amplitude of \% P-KaiC behaves near the phase boundary. We first focus on the transition at the standard KaiA concentration $c_\text{A}=1$, shown on the upper right of Fig.~\ref{fig_phase_transition}(a). As either $\mu$ or $\rho$ is tuned, the amplitude grows continuously from zero and scales as the square root of the distance to the phase boundary (see Fig.~\ref{fig_amplitude}(a) and Fig.~\ref{fig_amplitude}(b)), which is characteristic of a supercritical Hopf bifurcation \cite{strogatz2024nonlinear}.

Next, we examine the two phase transitions in Fig.~\ref{fig_KaiA}(b) at high or low $c_\text{A}$, when $\mu$ and $\rho$ are fixed. At high $c_\text{A}$, the amplitude similarly follows a square root dependence (see Fig.~\ref{fig_amplitude}(c)), which indicates a supercritical Hopf bifurcation. In contrast, the transition at low $c_\text{A}$ is qualitatively different (Fig.~\ref{fig_amplitude}(d)). As $c_\text{A}$ decreases from the synchronous phase, the unstable fixed point $\bx^\text{ss}$ becomes stable while the stable limit cycle persists. This suggests that the system undergoes a subcritical Hopf bifurcation, where the unstable fixed point splits into a stable fixed point and an unstable limit cycle. In this bistable regime, trajectories either approach the stable fixed point or the stable limit cycle depending on their initial conditions, which results in two branches for the amplitude (diamonds in Fig.~\ref{fig_amplitude}(d)). When $c_\text{A}$ is decreased further, the stable and unstable limit cycles annihilate in a saddle–node bifurcation of limit cycles, leaving only the stable fixed point $\bx^\text{ss}$. Beyond this point, the system no longer supports sustained oscillations, and trajectories only undergo damped oscillations toward $\bx^\text{ss}$.

\section{The diffusion approximation} \label{appx_diffusion}
In this section, we include a derivation of
the diffusion approximation, which converts a rate equation (Eq.~(\ref{eqn_coupled})) to a Langevin equation for more efficient simulations. Our derivation is similar to the approach outlined in Ref.~\cite{orio2012simple}.

We first set up the notation needed for the derivation. We consider a population of $N$ identical KaiC oscillators, each with $M$ possible states ($M=196$). The distribution of the population is specified by the population vector $\bx(t)=\frac{1}{N}(N_1(t),...,N_M(t))^\intercal$, where $N_i(t)$ is the number of molecules at state $i$. Transitions between system states are specified by the matrix (see Eq.~(\ref{eqn_coupled}))
\begin{equation}
    \cW(\bx) = \cW^\text{s} + c_\text{A}^\text{seq}(\bx)\cW',
\end{equation}
where $\cW_{ij}(\bx)$ denotes the transition rate from state $j$ to $i$. For notational simplicity, we suppress the explicit $\bx$-dependence in what follows. To enumerate all possible reactions, we introduce the $M\times K$ stoichiometric matrix $S$, where each column corresponds to one of the $K$ possible reactions. For the transition $j\rightarrow i$, the corresponding $k$-th column is given by
\begin{equation}
    S_{mk}=
    \begin{cases}
        +1, & m=i \\
        -1, & m=j \\
        0, & m\neq i, m\neq j
    \end{cases},
\end{equation}
which encodes the stoichiometric coefficient for the reaction.

The key idea of the diffusion approximation is to replace the Poisson processes describing biochemical reactions by Gaussian processes. 
Consider the number of reactions that occur within a short time interval $\pd t$. 
For the transition $j\rightarrow i$, the number of KaiC that 
undergo the transition follows a Poisson distribution with mean 
\begin{equation}
    \mu_{ij}(t)=\cW_{ij}N_j(t)\,\pd t=\cW_{ij}X_j(t)N\,\pd t +o(\pd t).
\end{equation}
Accounting for all possible reactions in the system, the total change of $N_i(t)$ over the interval $\pd t$ is 
\begin{equation} \label{eqn_dN}
    \pd N_i= \sum\limits_{j\neq i} \text{Poisson}[\mu_{ij}(t)] - \sum\limits_{j\neq i}\text{Poisson}[\mu_{ji}(t)]
\end{equation}

When $N$ is sufficiently large so that $\mu_{ij}(t)\gg 1$, the Poisson distributions may be approximated by Gaussian distributions with the same mean and variance $\mu_{ij}(t)$. Under this approximation, we can decompose the Gaussian increments into drift and diffusion terms, which yields a stochastic differential equation in the It\^o sense:
\begin{align}
    \pd N_i = &\left[\sum\limits_{j\neq i} \cW_{ij}X_j(t) - \sum\limits_{j\neq i} \cW_{ji}X_i(t)\right]N\,\pd t \nonumber\\ 
    &+ \sum\limits_{j\neq i} \sqrt{\cW_{ij}X_j(t)N} \,\pd W_{ij}(t) \nonumber\\
    &- \sum\limits_{j\neq i}\sqrt{\cW_{ji}X_i(t)N}\,\pd W_{ji}(t),
\end{align}
Here, $\pd W_{ij}(t)$ are  increments from independent Wiener processes. 
The stochastic differential equation is equivalent to the following Langevin equation:
\begin{align} \label{eqn_Xi}
    \frac{\pd X_i}{\pd t} = &\sum\limits_{j\neq i} \cW_{ij}X_j(t) - \sum\limits_{j\neq i} \cW_{ji}X_i(t) \nonumber\\ 
    &+ \frac{1}{\sqrt{N}}\sum\limits_{j\neq i} \sqrt{\cW_{ij}X_j(t)} \,\xi_{ij} \nonumber\\
    &- \frac{1}{\sqrt{N}}\sum\limits_{j\neq i} \sqrt{\cW_{ji}X_i(t)} \,\xi_{ji},
\end{align}
where $\xi_{ij}$ denote independent Gaussian white noises. 
Eq.~(\ref{eqn_Xi}) can be rewritten more compactly using the stoichiometric matrix. Let $j_k$ and $i_k$ denote the start and final state for the reaction corresponding to the $k$-th column of the stoichiometric matrix $S$. We then have
\begin{equation}
    \frac{\pd X_i}{\pd t} = \cW_{ij}X_j(t) + \frac{1}{\sqrt{N}}\sum\limits_{k}S_{ik}\sqrt{\cW_{i_k j_k}X_{j_k}(t)}\,\xi_{k}.
\end{equation}
Finally, we can write down the Langevin equation in vector form:
\begin{equation} \label{eqn_langevin2}
    \frac{\pd \bx}{\pd t}=\left[\cW^\text{s} + f(\bx)\cW'\right]\bx + \frac{1}{\sqrt{N}}G(\bx)\bm{\xi}(t),
\end{equation}
where $G(\bx)$ is an $M\times K$ matrix given by 
\begin{equation} \label{eqn_Gx}
    G_{ik}(\bx) = S_{ik}\sqrt{\cW_{i_k j_k}(\bx)X_{j_k}(t)}.
\end{equation}

In principle, converting the rate equation (Eq.~(\ref{eqn_coupled})) to the Langevin equation (Eq.~(\ref{eqn_langevin2})) requires that the mean $\mu_{ij}(t)=\cW_{ij}X_j(t)N\,\pd t\gg1$ for every transition $j\rightarrow i$ at any time $t$. Choosing $\text{d}t=10^{-3}T_1$, this condition demands very large populations on the order of $N\sim10^7$ for transitions with small rates (e.g., $\cW_{ij}=\gamma/e^{\mu+\rho}$). In practice, however, we find that much smaller values suffice: $N=500$ already leads to good agreement with Gillespie simulations as shown by Fig.~\ref{fig_phase_transition}(b) and Fig.~\ref{fig_phase_transition}(d). This likely reflects that reactions with small fluxes (where the approximation is the least precise) have minimal impact on the overall dynamics, allowing the diffusion approximation to be accurate at much smaller $N$.

One caveat of the approximation is that the Gaussian noise can drive the fractions $X_i(t)$ to negative values, which is unphysical. Several schemes have been proposed to address this issue \cite{orio2012simple,dangerfield2012modeling,pezo2014diffusion,pu2020fast}: (1) a free boundary condition where $X_i(t)$ is allowed to be negative; (2) a reflecting boundary condition where negative $X_i(t)$ is immediately reflected to $|X_i(t)|$; and (3) an absorbing boundary condition where negative values are set to 0. To compare their performance, we generated 500 trajectories of \% P-KaiC from each method and compared their average dynamics with those from Gillespie simulations, which we treat as the ground truth. The free boundary condition yields the best agreement, provided that we calculate the square root in Eq.~(\ref{eqn_Gx}) using the absolute value of the argument as done in Refs. \cite{orio2012simple,pu2020fast}. We therefore adopt the free boundary condition for all simulations in the main text involving the diffusion approximation, noting that negative excursions of $X_i(t)$ have small amplitudes and do not affect the dynamics noticeably.

\section{Extracting the timing error} \label{appx_timing_error}
To obtain the timing error, we extract the time between successive upward crossings of a fixed \% P–KaiC threshold, as illustrated on the top of Fig.~\ref{fig_phase_transition}(c). 
This method contrasts with Ref.~\cite{chew2018high}, which measures the oscillation period $T$ using peak-to-peak intervals. 
We use threshold crossings because it provides a more accurate estimate of the period: the uncertainty in determining the crossing time is inversely proportional to the derivative of \% P-KaiC at the threshold \cite{derenzo2014fundamental}. 
Since the derivative is small near the peaks, small fluctuations in \% P-KaiC produce much more jitters in peak-to-peak intervals compared to a threshold where the derivative is large.

To study the scaling behavior of the timing error, we generate at least 1000 trajectories at each $N$. Each trajectory is smoothed using the Savitsky-Golay filter, again using third-order polynomial fitting and a window size of 61 time points. 
The period $T$ is then extracted from the points at which the smoothed trajectories increase past the threshold \% P-KaiC level. Note that we smooth out individual trajectories here and not the averaged trajectory, in contrast to Appendix~\ref{appx_coherence}. The choice of threshold depends on the parameter regime. 
For Fig.~\ref{fig_phase_transition}(d) where $c_\text{A}=1$, we choose a threshold of 0.62 for all $N$ and $(\mu,\rho)$, the level at which \% P-KaiC changes rapidly. 
In contrast, for Fig.~\ref{fig_KaiA}(d), the mean level of \% P-KaiC oscillations shifts significantly with KaiA concentration. 
We thus select a separate threshold for each combination of $(\mu,\rho,c_\text{A})$ by minimizing the resulting timing error. 
In practice, this optimal threshold lies around the midpoint between the peak and the trough of the average oscillation profile, where spurious crossings from random fluctuations (which increases timing error) are the least likely. 
After extracting the crossing times and therefore the periods $T$, the timing error is computed from the coefficient of variation for $T$, i.e., $\text{Std}[T]/\text{E}[T]$.

\begin{figure}[htb!]
	\centering
	\includegraphics[width=8cm, height=5.8cm]{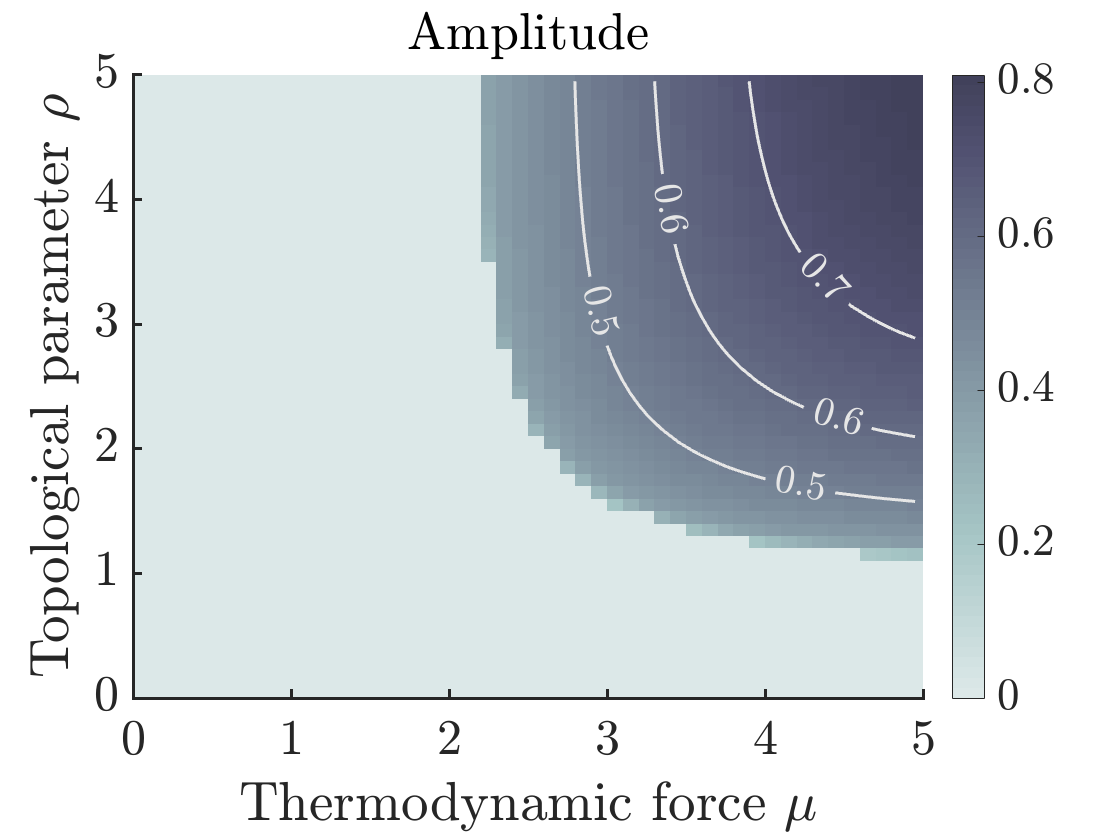}
	\caption{Oscillation amplitude increases with both the thermodynamic force $\mu$ and topological parameter $\rho$. The amplitude is set by the single-oscillator coherence $\cR_1$: parameters $(\mu,\rho)$ with the same $\cR_1$ yield roughly the same amplitude.}
    \label{fig_amplitude_full}
\end{figure} 

\section{Single-oscillator coherence $\cR_1$ and amplitude} \label{appx_R1_amp}

In our model, larger single-oscillator coherence $\cR_1$ is associated with a larger amplitude for population-level \% P-KaiC oscillations. 
This effect is shown in Fig.~\ref{fig_amplitude_full}: the amplitude increases with both the thermodynamic force $\mu$ and the topological parameter $\rho$, and its contour lines closely follow that of $\cR_1$ in Fig. \ref{fig_phase_transition}(a). This structure indicates that parameters $(\mu,\rho)$ with the same $\cR_1$ gives rise to similar amplitudes.

This connection arises because more coherent oscillations on the single-molecule level allow the population to be more synchronized: individual oscillators are less susceptible to stochastic fluctuations that spread out their phosphorylation states, leading to a narrower distribution throughout the oscillation cycle. In our model, individual oscillators
typically traverse the boundary of the state space, due to topological protection, meaning that they become fully phosphorylated or dephosphorylated within each cycle. As a result, the narrower distribution translates into higher peaks (more molecules at full phosphorylation) and lower troughs (more molecules at full dephosphorylation), and hence a larger amplitude. This connection between $\cR_1$ and amplitude explains why longer oscillation periods (increasing $\rho$) and stronger non-equilibrium driving (increasing $\mu$) are associated with a larger oscillation amplitude in our model.

\section{Estimating the thermodynamic force $\mu$ from experiments} \label{appx_estimate}
In this section, we give an estimate of the thermodynamic force $\mu$ in our model based on experimentally measured ATP consumption rates for KaiC. 
This estimate allows us to identify the experimentally relevant parameter regime as shown in Fig. \ref{fig_KaiA}(b). 

In our model, $\mu$ is defined as the log ratio of forward to reverse rates for each pair of transitions. 
Based on non-equilibrium thermodynamics, this quantity is set by the free energy input from ATP hydrolysis, which maintains the KaiABC circadian oscillations \cite{hill1989}. 
For simplicity, we assume the same $\mu$ for each pair of reactions in the state space. 

\textit{In vitro} experiments have found that each KaiC monomer consumes around 12--15 ATP molecules per day (i.e., per oscillation cycle) \cite{terauchi2007atpase,chavan2021reconstitution}. 
Assuming that hydrolysis of each ATP molecule releases 20 $k_\text{B}T$ of free energy \cite{milo2015cell}, each KaiC hexamer dissipates 1400--1800 $k_\text{B}T$ over each cycle. 
Ref.~\cite{zheng2024topological} has demonstrated that such dissipation is not only localized on the boundary of state space: the steady-state dissipation rates for reactions in the bulk are comparable to those on the boundary. 
We thus assume that the free energy dissipated per oscillation cycle is divided equally among all $364$ pairs of forward/reverse reactions in our state space. 
This gives a thermodynamic force ranging from 3.8 to 4.9 $k_\text{B}T$.

The above estimate corresponds to standard \textit{in vitro} experimental conditions \cite{kageyama2006cyanobacterial,nishiwaki2007sequential}. 
To map out the full experimentally relevant region in Fig.~\ref{fig_KaiA}(b), we also consider experiments with reduced ATP levels. 
For example, Phong et al.~2013 has observed sustained KaiC oscillations for ATP fractions down to 25\% \cite{phong2013robust}. 
Under these conditions, we estimate the free energy released from ATP hydrolysis by
\begin{equation}
    \Delta G = \Delta G ^0 + \ln{\frac{[\text{ATP}]}{[\text{ADP}][\text{P}_\text{i}]}},
\end{equation}
where $\Delta G ^0\approx 12\;k_\text{B} T$ \cite{milo2015cell}. With $[\text{ATP}]/[\text{ADP}]=1/3$ at 25\% ATP and assuming trace amounts of $\text{P}_\text{i}$ present in solution, hydrolysis can still release $\sim15$ $k_\text{B}T$ of free energy per molecule. This leads to a thermodynamic force of $\sim3\;k_\text{B}T$, included in the hatched region of Fig. \ref{fig_KaiA}(b). 

\end{appendices}

\bibliography{references}

\end{document}